\documentclass[conference]{IEEEtran}
\usepackage{siunitx}
\usepackage{color}

\ifCLASSINFOpdf
  \usepackage[pdftex]{graphicx}
  \graphicspath{{./Figures/}}
  \DeclareGraphicsExtensions{.pdf,.jpeg,.jpg,.png}
\else
\fi
\usepackage{array}

\ifCLASSOPTIONcompsoc
 \usepackage[caption=false,font=normalsize,labelfont=sf,textfont=sf]{subfig}
\else
 \usepackage[caption=false,font=footnotesize]{subfig}
\fi
\definecolor{orange}{rgb}{1, .36, .08}
\definecolor{pink}{rgb}{1.00, 0.00, 0.5}

\DeclareSIUnit\mAh{mAh}

\renewcommand{\baselinestretch}{0.995}

\begin{document}
%

\title{
An EMG Gesture Recognition System with\\ 
Flexible High-Density Sensors and\\ 
Brain-Inspired High-Dimensional Classifier
}


%
\author{\IEEEauthorblockN{Ali Moin\IEEEauthorrefmark{1}, Andy Zhou\IEEEauthorrefmark{1}, Abbas Rahimi\IEEEauthorrefmark{1}\IEEEauthorrefmark{2}, Simone Benatti\IEEEauthorrefmark{3}, Alisha Menon\IEEEauthorrefmark{1}, Senam Tamakloe\IEEEauthorrefmark{1}, Jonathan Ting\IEEEauthorrefmark{1},\\
Natasha Yamamoto\IEEEauthorrefmark{1}, Yasser Khan\IEEEauthorrefmark{1}, Fred Burghardt\IEEEauthorrefmark{1}, Luca Benini\IEEEauthorrefmark{2}\IEEEauthorrefmark{3}, Ana C. Arias\IEEEauthorrefmark{1}, Jan M. Rabaey\IEEEauthorrefmark{1}
}
\IEEEauthorblockA{\IEEEauthorrefmark{1}Berkeley Wireless Research Center, EECS Department, 
University of California, Berkeley.}
\IEEEauthorblockA{\IEEEauthorrefmark{2}Integrated System Laboratory, ETH Zurich, Switzerland. \IEEEauthorrefmark{3}DEI, University of Bologna, Italy.}
\IEEEauthorblockA{Corresponding Author Email: moin@berkeley.edu}
}


\maketitle

\begin{abstract}
EMG-based gesture recognition shows promise for human--machine interaction. Systems are often afflicted by signal and electrode variability which degrades performance over time.
We present an end-to-end system combating this variability using a large-area, high-density sensor array and a robust classification algorithm. EMG electrodes are fabricated on a flexible substrate and interfaced to a custom wireless device for 64-channel signal acquisition and streaming.
We use brain-inspired high-dimensional (HD) computing for processing EMG features in one-shot learning. The HD algorithm is tolerant to noise and electrode misplacement and can quickly learn from few gestures without  gradient descent or back-propagation.
We achieve an average classification accuracy of 96.64\% for five gestures, with only 7\% degradation when training and testing across different days. 
Our system maintains this accuracy when trained with only three trials of gestures; it also demonstrates comparable accuracy with the state-of-the-art when trained with one trial.
\end{abstract}


%
\IEEEpeerreviewmaketitle


\section{Introduction}
Hand gestures are an integral part of human--human communication and can be leveraged for more natural human--machine interaction (HMI). They are a fast and effective physical medium for communicating with and controlling intelligent devices. Accurate and efficient gesture recognition enables intuitive control for applications ranging from polyarticulated prosthetic hands~\cite{benatti2017prosthetic} to mobile and game interfaces~\cite{zhang2009hand}. Hand gestures may be recognized based on muscular activity measured using electromyography (EMG), the detection of field potentials representing the superimposed electrical activity of muscle fibers. Surface EMG is a non-invasive method of acquiring these signals by placing recording electrodes directly on the surface of the skin.

Both research work~\cite{lda,ann,svm} and commercial products~\cite{myoarmband} have demonstrated great potential for machine learning techniques to decode EMG and enable natural gesture recognition. Most of these systems are composed of an array of EMG sensors placed on the forearm and connected to a PC for acquiring data and running pattern recognition algorithms~\cite{benatti2017prosthetic}. Over a single acquisition session, these systems can achieve classification accuracies above 90\% in recognition of 5-6 gestures with an array of 3-8 sensors. Nevertheless, classification accuracy depends highly on precise electrode positioning for sufficient muscle coverage. For systems with low electrode-counts, inaccurate positioning can result in recording insufficient information for gesture recognition. 

Furthermore, the EMG signal can be highly variable, and it is challenging to have a reliable interface which reaches the same accuracy over multiple acquisition sessions. This performance variability is mostly caused by muscle fiber crosstalk, skin perspiration, and by changes in the skin-electrode interface. Moreover, even small variations in sensor positioning over multiple sessions can further decrease the classification performance. For instance, an accuracy degradation of 27\% is observed when training and testing are performed on different sessions for a 14-sensor EMG system classifying 7 gestures using a random forest algorithm~\cite{palermo2017repeatability}.
In the same vein, a support vector machine (SVM) algorithm suffers from a 30\% accuracy drop with a single training session and multiple classification sessions~\cite{biosignals14}.

A promising solution to assure sufficient muscle coverage is to acquire EMG signals from a dense array of sensors~\cite{otbioe}, obtaining a fine-grained coverage of the muscular surface.
Commercial active sensors~\cite{Ottobock} are not suited for use with dense arrays as they are large and cumbersome due to the integrated active conditioning circuitry.
Passive single-lead electrodes must be applied one-by-one, which is inconvenient for daily use of the HMI.
In contrast, a matrix of passive EMG sensors can be built into a flexible printed circuit board (PCB) and easily positioned on the skin, enabling high-density and small form-factor multichannel EMG acquisition~\cite{otbioe}. All electrodes can be placed at once, and the flexible nature of such an array allows it to conform to the curvature of the arm, ensuring good signal quality. Dense electrode spacing on a large array increases channel-count and can easily lead to bulky signal acquisition systems, so a small form-factor and wireless device is desirable. 

Additionally, an intrinsically robust classification algorithm is needed to process the large number of noisy inputs and mitigate the effects of signal variability. Brain-inspired high-dimensional (HD) computing~\cite{HD09} is a promising avenue that can overcome low signal-to-noise ratio and large data variability to perform robust decision making and classification~\cite{HDlowSNR}.
The HD computational paradigm provides generality, scalability, and fast one-shot learning, making it a prime candidate for processing
multidimensional sensor data such as EMG~\cite{rahimi2016hyperdimensional}, electroencephalography (EEG)~\cite{HD_EEG}, etc.

This paper presents two major contributions: (1) an end-to-end, high-density, wireless EMG gesture recognition system featuring a compact wearable device; and (2) the use of brain-inspired HD computing to improve the robustness of classification using variable data from a large number of noisy sensors. The proposed system wirelessly acquires 64 channels of EMG from a flexible PCB sensor array connected to a dedicated biopotential readout device~\cite{omni,wand}. 
We implement an enhanced version of our previous work demonstrating HD computing for classification of EMG signals recorded from four gel-based electrodes~\cite{rahimi2016hyperdimensional}.
We extensively evaluate the classification accuracy and its robustness over three acquisition sessions, including repositioning of the array strip.
We show how the HD algorithm is easily scaled to operate with 64 channels, tolerates noise, and can be trained quickly from a single trial of each gesture, paving the way toward extremely fast calibration and online learning. 

\begin{figure}[t]
	\centering
	\subfloat{\includegraphics[width=0.5\columnwidth]{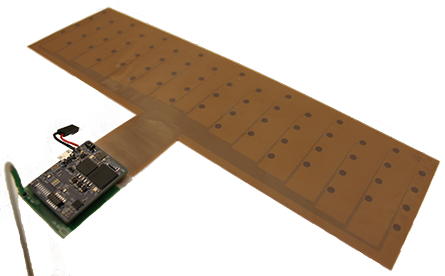}}
    \hfil
    \subfloat{\includegraphics[width=0.5\columnwidth]{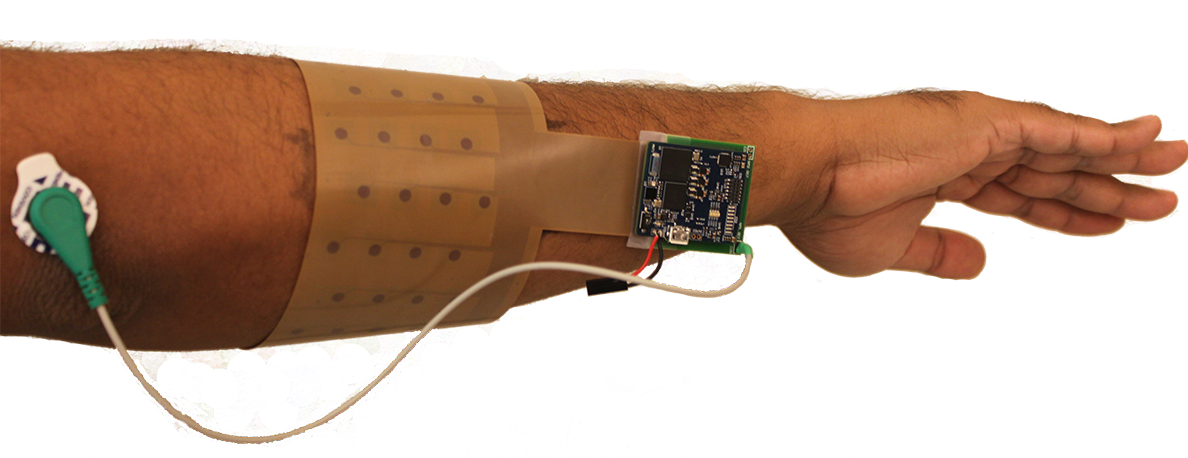}}
	\caption{Flexible electrode array with attached wireless biosignal acquisition device (left) positioned on the arm (right).}
	\label{fig:array}
\end{figure}

\section{System Overview}
The gesture recognition device is composed of two main components: a high-density flexible electrode array, and a wireless neural-signal acquisition device.


\subsection{Flexible Electrode Array}
A high-density flexible electrode array serves as the interface between the skin and the neural recording circuitry (Figure~\ref{fig:array}). 64 uniformly-distributed electrodes are laid out in a 16 $\times$ 4 grid on a \SI{200}{\um}-thick flexible substrate wide enough to cover the circumference of the forearm (\SI{29.3}{\cm} $\times$ \SI{8.2}{cm}). The size and flexibility of the substrate guarantee full, stable coverage of all muscles used for different gestures. The electrodes (\SI{4.3}{\mm} diameter) and traces are fabricated out of copper. A piece of conductive hydrogel tape is applied to each electrode to help maintain good contact with the skin. 

Interfacing flexible electronics with rigid PCBs (recording module) is often challenging. This is solved by designing a small form-factor adapter board which has a zero insertion force (ZIF) connector for connecting the flexible array on one side, and a DF-12 connector for connecting the recording module on the other side.



\subsection{Wireless Neural Recording Module}

\begin{figure}[t]
	\centering
	\includegraphics[width=0.9\columnwidth]{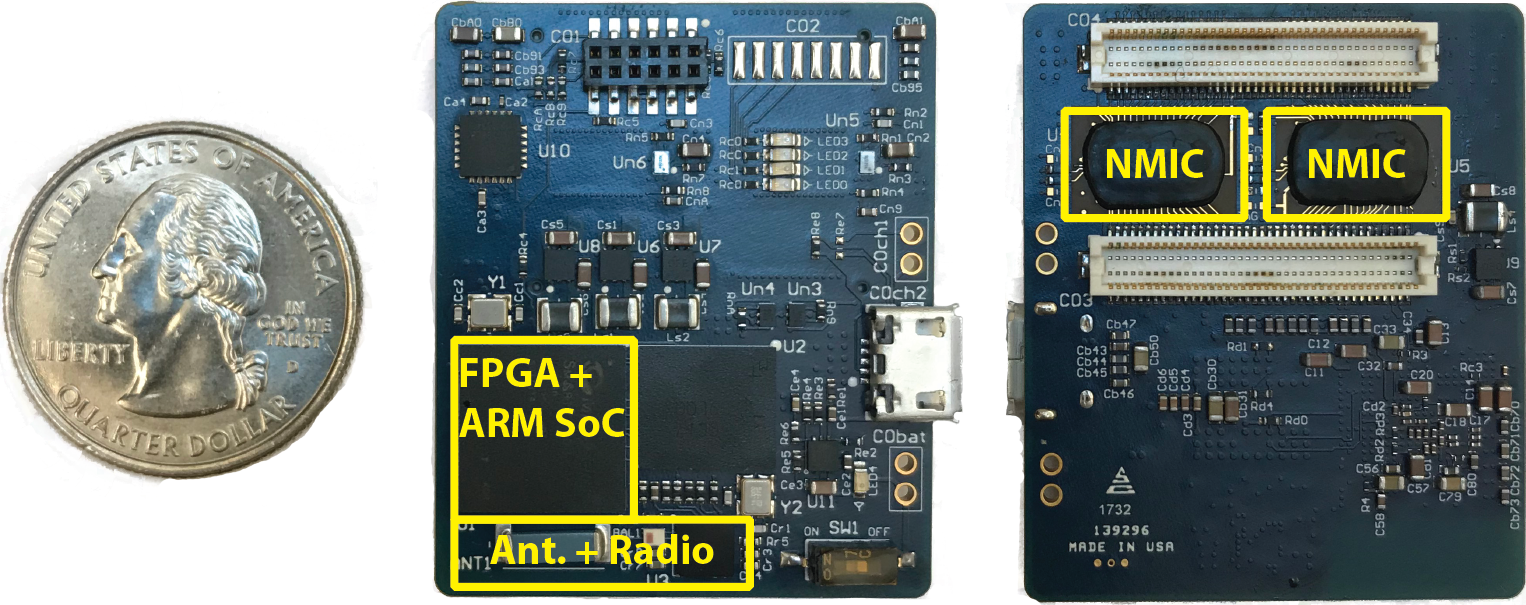}
	\caption{Wireless biosignal acquisition device with important blocks annotated.}
	\label{fig:module}
\end{figure}

A compact wireless module attaches to and interfaces with the electrode array to record, digitize, and wirelessly transmit the raw EMG signals to a base station (Figure~\ref{fig:module}). The device is based on our previous design for a closed-loop artifact-free neuromodulation platform~\cite{omni,wand}. Two custom neuromodulation ICs~\cite{nmic} (NMICs by Cortera Neurotechnologies, Inc.) provide 128 recording front-ends in a small footprint, and an SoC FPGA with a \SI{166}{\MHz} ARM Cortex-M3 processor (SmartFusion2 M2S060T, Microsemi) acts as the master module aggregating data. The digitized raw data is transmitted by a \SI{2.4}{\GHz} low-energy radio (nRF51822, Nordic Semiconductor) to a base-station. Table~\ref{tab:specs} summarizes main device specifications. The device is powered by a \SI{500}{\mAh} \SI{4.1}{\V} Li-ion battery providing up to 11 hours of streaming.

\begin{table}[!t]
\caption{Wireless Biosignal Acquisition Device Specifications}
\label{tab:specs}
\centering
\begin{tabular}{|c|c|} \hline
Number of Recording Channels & 128\\ \hline
ADC Sample Rate & 1 \si{\kilo S\per\second}\\ \hline
ADC Resolution & 15 bits\\ \hline
Input Range & \SI{100}{\mV pp}\\ \hline
Noise Floor & \SI{1.65}{\uV}rms\\ \hline
Number of Channels Wirelessly Streamed & 96\\ \hline
Wireless Data Rate & \SI{2}{\mega bp \second}\\ \hline
PCB Dimensions & \SI{3.3}{\cm} $\times$ \SI{3.56}{\cm}  \\ \hline
Weight (w/ battery) & \SI{6.62}{\gram} (\SI{17.18}{\gram})  \\ \hline
Battery Life & \SI{11}{hr} \\ \hline
\end{tabular}
\end{table}

\begin{figure*}[!t]
	\centering
	\includegraphics[width=0.9\textwidth]{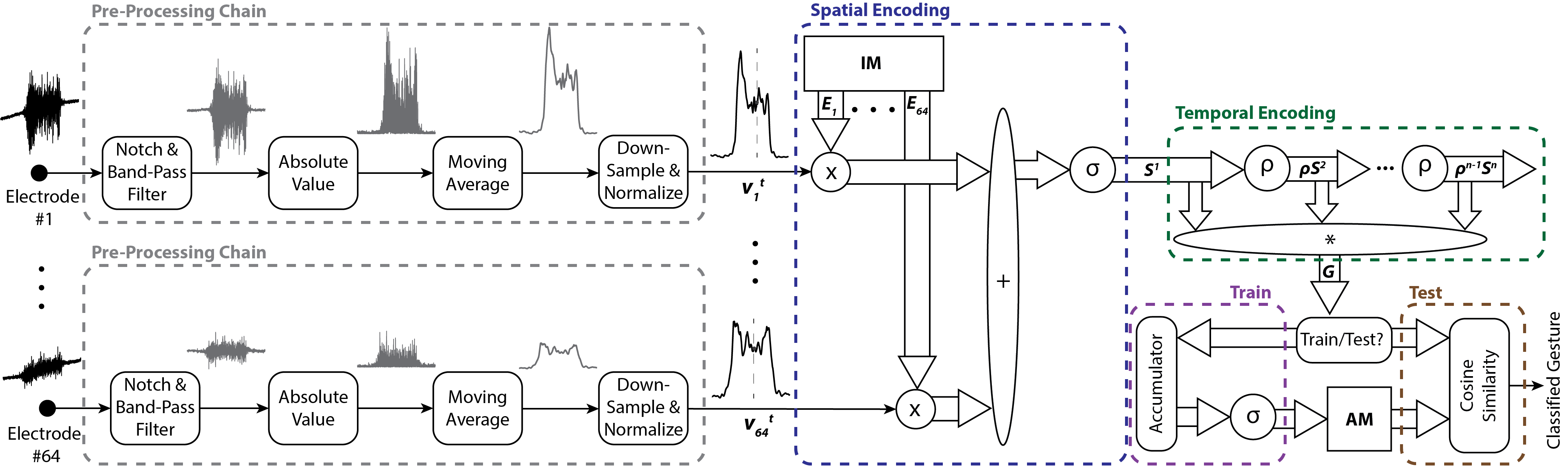}%
	\caption{Preprocessing and spatiotemporal encoding for EMG-based gesture recognition where v$_i^t$ is the preprocessed scalar value of electrode $i$ at time $t$, $S^t$ (spatial HD vector) is the output of spatial encoder at time $t$, and $G$ (spatiotemporal HD vector) is the output of temporal encoder.}
  	\vspace{-10pt}
	\label{fig:hd}
\end{figure*}

Data is streamed to a wireless base-station connected to a laptop running a text-based MATLAB (MathWorks, Inc.) GUI for configuring the device, instructing the subject, visualizing and storing data, and performing classification.

\section{High-dimensional (HD) Computing}
The human brain contains billions of neurons and synapses, suggesting that large circuits are fundamental to its computational power.
High-dimensional (HD) computing~\cite{HD09} explores this idea by looking at computing with vectors of very high dimensionality, e.g. 10,000.
HD vectors can be combined into new vectors using well-defined vector space operations while maintaining the original information with high probability.
They can be compared for similarity using a distance metric over the HD vector space.

HD vectors are initially taken from a 10,000-dimensional space and have an equal number of randomly placed $+1$s and $-1$s.
Such HD vectors are used to represent the basic elements in the system (e.g., the electrodes~\cite{rahimi2016hyperdimensional,HD_EEG}), and are stored in an item memory (IM) that functions as a fixed symbol table. 

The following vector space operations can be used on the elements of the IM to encode information:
Point-wise multiplication (*), or binding, takes two vectors and yields a third vector that is dissimilar (orthogonal) to the two. Point-wise addition (+), or bundling, takes several vectors and yields their mean vector that is maximally similar to all of them.
These operations, along with scalar multiplication ($\times$) and permutation ($\rho$) of vector components for sequences, form an algebraic field beyond arithmetic and linear algebra. 

\subsection{Preprocessing}
The EMG signal is recorded from 64 single-ended channels all referenced to a single Ag/AgCl patch electrode placed on the elbow. The acquired signal is a mixture of the EMG potentials and time-varying offset and noise, but the desired feature is the envelope of the high-frequency EMG. Hence, some preprocessing is needed in order to make it suitable for the HD classifier. 

The preprocessing chain is illustrated in Figure~\ref{fig:hd}. The first step is the elimination of the power line interference by a notch filter at \SI{60}{Hz} with Q-factor of 50. Additionally, an 8th-order Butterworth band-pass filter for frequencies between \SI{1} and \SI{200}{Hz} cancels the undesired frequency components such as DC offset and drift.

To extract the envelope, we take the absolute value of the signal and apply a moving average filter with window size of 100. 
Finally, the data is normalized per channel, and down-sampled by a factor of 100, i.e. 10 samples per second are fed to the HD classifier described in the next section.

\subsection{HD classifier}
The HD algorithm encodes windows of EMG signals into HD vectors that are ultimately used for robust learning and classification.
The input features at each time point are the preprocessed and downsampled scalar values for each electrode. We encode the input features at a single time point into a \emph{spatial} HD vector, $S^t$.
The IM assigns a unique orthogonal HD vector to every electrode, i.e., $E_1 \perp E_2 ... \perp E_{64}$.
To represent the preprocessed scalar value ($v_i^t$) of an electrode $i$ at time $t$, we simply multiply the scalar with the corresponding HD vector: $E_i \times v_i^t$.
These vectors are added across all electrodes to compute the spatial HD vector representing the input features: $S^t=\sigma(\sum_{i=1}^{64} E_i \times v_i^t)$ where $\sigma$ is a bipolar thresholder that turns a positive element to $+1$ and a negative element to $-1$.
This new spatial encoder computes the sum of the electrode vectors weighted by the scalars, and \emph{naturally} maps a large number of features to a single spatial HD vector.

Next, we need to encode a sequence of $n$ spatial HD vectors to capture relevant temporal information into a final spatiotemporal vector by using permutation: $G=\prod_{t=1}^{n} \rho^{t-1} S^t$ where $\rho^{k}$ is a rotation over $k$ positions of the HD vector. For the preprocessed EMG signals subsampled at 10\,Hz, we observe that a temporal window of 5 samples, i.e., $n=5$ yields the best classification results.

To train the classifier, we use the $G$ spatiotemporal HD vectors to build an associative memory (AM) containing an HD vector representing each gesture label. 
During training, all spatiotemporal HD vectors computed over a labeled gesture window are accumulated (summed) to form a \emph{prototype} HD vector representing that gesture. 
The prototype HD vector is thresholded by $\sigma$ and stored in the AM. 
For classification, newly computed spatiotemporal HD vectors are compared to each entry of the AM using cosine similarity as the distance metric. The classified gesture is that of the closest gesture vector in the AM.




\section{Experimental Results}
\subsection{Generating Dataset}
We acquired EMG gesture data from three healthy, adult male subjects\footnote{Dataset and processing scripts available at https://github.com/a-moin/flexemg}. 
Each subject participated in three data collection sessions across different days, where the electrode array was reapplied with fresh hydrogel tape for each session. For Sessions 1 and 2, the array was placed in approximately the same position each time (Figure~\ref{fig:array}). A training data set and testing data set were recorded 30 minutes apart during Session 1, and a second testing data set was recorded 24 hours later in Session 2. For Session 3, new training and testing data sets were recorded with the array in a different orientation 1 week after Sessions 1 and 2. 

\begin{figure}[t]
	\centering
	\includegraphics[width=0.9\columnwidth]{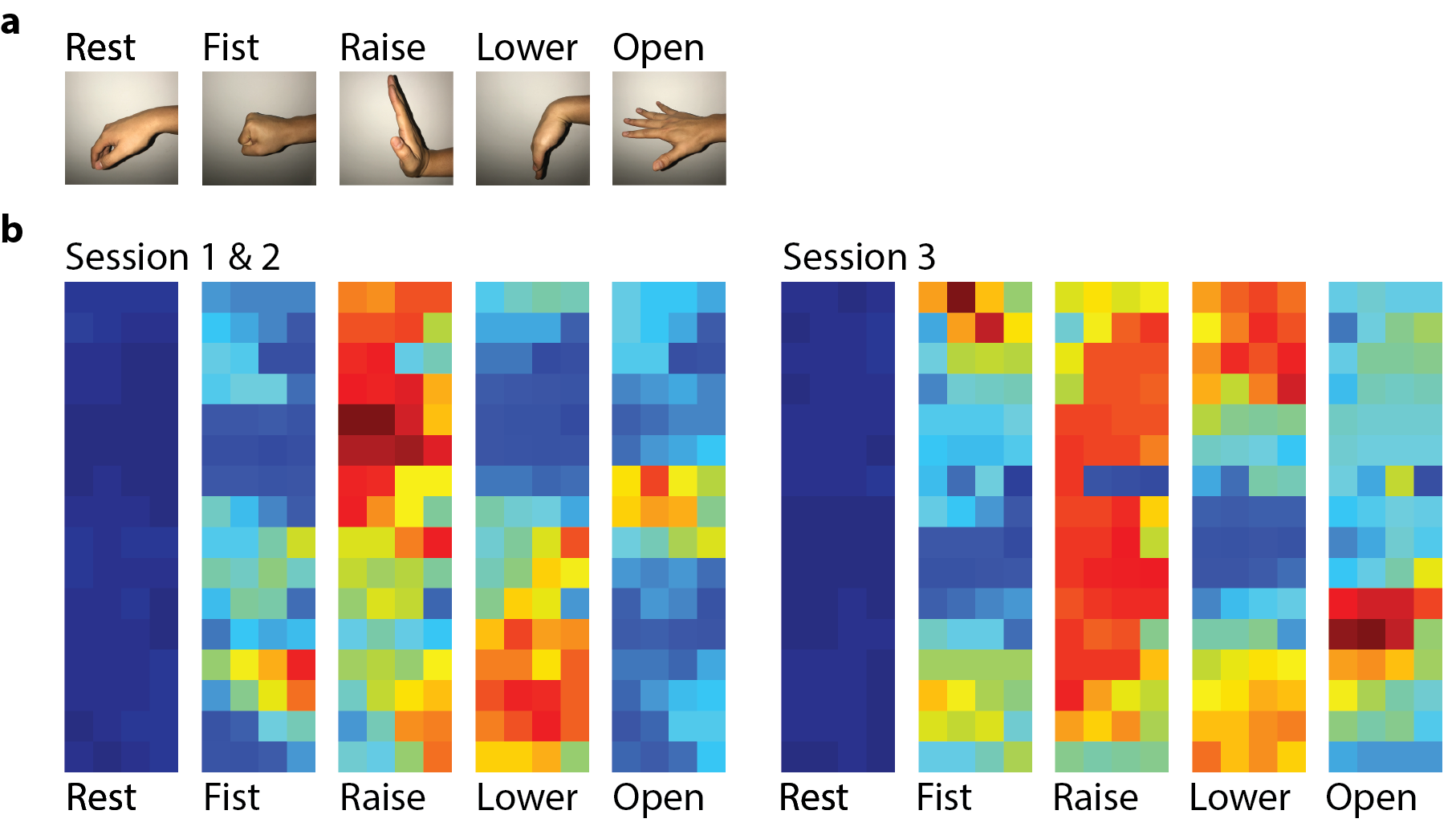}
	\caption{Dataset gestures (a) with the associated normalized activity maps for different sessions (Subject 1 shown) (b). Pixels correspond to electrode positions in the array.}
	\label{fig:gestures}
\end{figure}

Each data set contained ten trials of four hard gestures (fist, raise, lower, open) (Figure~\ref{fig:gestures}) held for 5 seconds each in different sequences. Each sequence began and ended with rest, which we treated as a fifth gesture. Centered 3-second segments of each 5-second gesture were labeled for inclusion in the testing and training sets. 

Preprocessed features for each gesture were averaged and arranged in matrices for visualization as heat maps of muscular activity (Figure~\ref{fig:gestures}). For Session 3, the array was rotated from its Session 1-2 position. This can be seen in the differences of the heat maps, though the large, high-density array maintained full coverage of muscle activity across all sessions. 

\subsection{Classification Results}

\begin{table}[!t]
\scriptsize
\caption{Classification Accuracy Results}
\label{tab:accs}
\centering
\begin{tabular}{|c|c|c|c|c|c|c|}
\hline
& \multicolumn{2}{c|}{\textbf{Same Session}} & \multicolumn{2}{c|}{\textbf{Across Sessions}} & \multicolumn{2}{c|}{\textbf{Same Session Rotated}} \\ \cline{2-7}
& \textbf{No Vote} & \textbf{1s Vote} & \textbf{No Vote} & \textbf{1s Vote} & \textbf{No Vote} & \textbf{1s Vote} \\ \hline
\textbf{Sub. 1} & 99.44\% & 100\% & 90.97\% & 93.26\% & 99.57\% & 100\% \\ \hline
\textbf{Sub. 2} & 98.87\% & 98.99\% & 98.68\% & 98.88\% & 97.67\% & 98.29\% \\ \hline
\textbf{Sub. 3} & 91.61\% & 93.03\% & 79.69\% & 82.64\% & 90.81\% & 95.61\% \\ \hline
\textbf{Avg.} & \textbf{96.64}\% & \textbf{97.34}\% & \textbf{89.78}\% & \textbf{91.59}\% & \textbf{96.02}\% & \textbf{97.97}\% \\ \hline
\end{tabular}
\end{table}

\begin{figure}[t]
\centering
	\includegraphics[width=0.9\columnwidth]{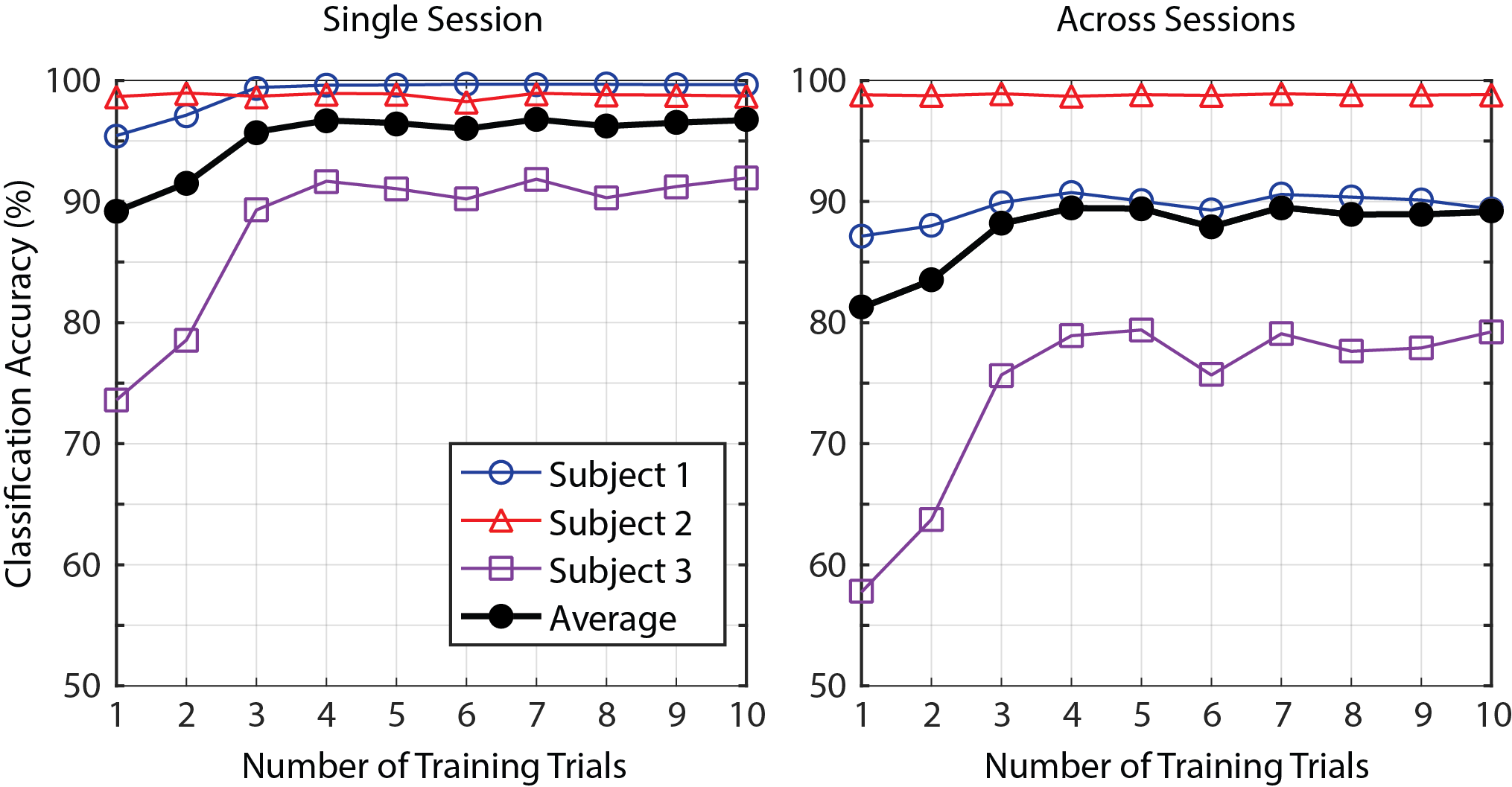}
	\caption{Classification accuracy for different number of training trials.}
	\vspace{-15pt}
	\label{fig:numtrials}
\end{figure}

To quantify single- and across-session classification accuracies, the HD classifier was trained on Session 1 and tested on Sessions 1 and 2. During testing, we generated a classification result for every 500-ms window sliding by 100 ms. Accuracies for three subjects, calculated as the percentage of classification results that matched the labeled gesture, are shown in Table~\ref{tab:accs}.  Single-session accuracies were on average 96.64\%. For comparison, single-session accuracy using an SVM classifier on similarly processed signals~\cite{Ottobock} is 88.96\% for 5 gestures~\cite{biosignals14}. Training and testing across Sessions 1 and 2 resulted in accuracy degradation of only about $7\%$, a large improvement over degradations of more than $30\%$ using the SVM classifier~\cite{biosignals14}. We attribute this robustness to the high channel density and ability of HD computing to overcome variance in signal quality and electrode placement.

Simple majority voting of the classification results over a 1.1-second window of classifications (11 results) further improved accuracy on average by $1.5\%$. This is an insignificant improvement due to high baseline accuracy.

The HD algorithm was also successfully used to train and test on Session 3, demonstrating that the high-density array could record sufficient information for accurate classification in multiple orientations. When trained on Session 1 data, the classifier could not classify Session 3 gestures directly with good accuracy. However, we anticipate that a simple remapping of the input features could be implemented to recover the accuracy.

Limiting the number of trials used for training the classifier only marginally degraded performance. Figure~\ref{fig:numtrials} plots the classification accuracy averaged over all three subjects for training set sizes from 1 to 10 trials. Maximum performance ($96.64\%$) is achieved when training on only 3 trials, and training on a single trial can deliver acceptable accuracy ($89.19\%$) for true one-shot learning.

\section{Conclusion}
Robustness and reliability of gesture recognition are big challenges in designing an EMG-based HMI. Electrode misplacement over multiple sessions is the root cause of accuracy degradation, which can be higher than 30\%. We have presented a system for EMG-based gesture classification combining a flexible high density electrode array, a dedicated biopotential acquisition device, and a brain-inspired classification algorithm. Large area coverage and dense electrode spacing ensures sufficient muscular coverage without requiring precise placement. Furthermore, the wireless and compact signal acquisition device promotes comfort and ease of use. Finally, the HD algorithm achieves high classification accuracy without substantial degradation over multiple sessions, and can be trained using minimal amounts of data. 



\section*{Acknowledgment}
The authors would like to thank George Alexandrov, Ken Lutz, Profs. Elad Alon and Rikky Muller, and Cortera Neurotechnologies Inc. This work was supported in part by Semiconductor Research Corporation (SRC) under the STARnet SONICS and TerraSwarm centers and the JUMP CONIX center. Support was also received from sponsors of Berkeley Wireless Research Center, NSF Graduate Research Fellowship under Grant No. 1106400, ETH Zurich Postdoctoral Fellowship program and the Marie Curie Actions for People COFUND Program.



\bibliographystyle{IEEEtran}
\bibliography{refs}

\begin{thebibliography}{10}
\providecommand{\url}[1]{#1}
\csname url@samestyle\endcsname
\providecommand{\newblock}{\relax}
\providecommand{\bibinfo}[2]{#2}
\providecommand{\BIBentrySTDinterwordspacing}{\spaceskip=0pt\relax}
\providecommand{\BIBentryALTinterwordstretchfactor}{4}
\providecommand{\BIBentryALTinterwordspacing}{\spaceskip=\fontdimen2\font plus
\BIBentryALTinterwordstretchfactor\fontdimen3\font minus
  \fontdimen4\font\relax}
\providecommand{\BIBforeignlanguage}[2]{{%
\expandafter\ifx\csname l@#1\endcsname\relax
\typeout{** WARNING: IEEEtran.bst: No hyphenation pattern has been}%
\typeout{** loaded for the language `#1'. Using the pattern for}%
\typeout{** the default language instead.}%
\else
\language=\csname l@#1\endcsname
\fi
#2}}
\providecommand{\BIBdecl}{\relax}
\BIBdecl

\bibitem{benatti2017prosthetic}
S.~Benatti, B.~Milosevic, E.~Farella, E.~Gruppioni, and L.~Benini, ``A
  prosthetic hand body area controller based on efficient pattern recognition
  control strategies,'' \emph{Sensors}, vol.~17, no.~4, p. 869, 2017.

\bibitem{zhang2009hand}
X.~Zhang, X.~Chen, W.-h. Wang, J.-h. Yang, V.~Lantz, and K.-q. Wang, ``Hand
  gesture recognition and virtual game control based on {3D} accelerometer and
  {EMG} sensors,'' in \emph{Proceedings of the 14th international conference on
  Intelligent user interfaces}.\hskip 1em plus 0.5em minus 0.4em\relax ACM,
  2009, pp. 401--406.

\bibitem{lda}
H.~Zhang, Y.~Zhao, F.~Yao, L.~Xu, P.~Shang, and G.~Li, ``An adaptation strategy
  of using lda classifier for {EMG} pattern recognition,'' in \emph{EMBC, 2013
  35th Annual International Conference of the IEEE}, July 2013.

\bibitem{ann}
M.~Ahsan, M.~Ibrahimy, and O.~Khalifa, ``Electromygraphy ({EMG}) signal based
  hand gesture recognition using artificial neural network (ann),'' in
  \emph{Mechatronics (ICOM), 2011 4th International Conference On}, May 2011,
  pp. 1--6.

\bibitem{svm}
M.~A. Oskoei and H.~Hu, ``Support vector machine-based classification scheme
  for myoelectric control applied to upper limb,'' \emph{IEEE transactions on
  biomedical engineering}, vol.~55, no.~8, pp. 1956--1965, 2008.

\bibitem{myoarmband}
\BIBentryALTinterwordspacing
``{Thalmic Labs Myo Armband},'' 2017, (Date last accessed 20-Nov-2017).
  [Online]. Available: \url{https://www.thalmic.com/}
\BIBentrySTDinterwordspacing

\bibitem{palermo2017repeatability}
F.~Palermo, M.~Cognolato, A.~Gijsberts, H.~M{\"u}ller, B.~Caputo, and
  M.~Atzori, ``Repeatability of grasp recognition for robotic hand prosthesis
  control based on {sEMG} data,'' in \emph{Rehabilitation Robotics (ICORR),
  2017 International Conference on}.\hskip 1em plus 0.5em minus 0.4em\relax
  IEEE, 2017, pp. 1154--1159.

\bibitem{biosignals14}
S.~Benatti, E.~Farella, E.~Gruppioni, and L.~Benini, ``Analysis of robust
  implementation of an {EMG} pattern recognition based control,'' in
  \emph{Proceedings of the International Conference on Bio-inspired Systems and
  Signal Processing (BIOSTEC 2014)}, 2014, pp. 45--54.

\bibitem{otbioe}
\BIBentryALTinterwordspacing
``{OT Bioelettronica Electrodes},'' 2014, (Date last accessed 20-Nov-2017).
  [Online]. Available: \url{http://www.otbioelettronica.it/}
\BIBentrySTDinterwordspacing

\bibitem{Ottobock}
\BIBentryALTinterwordspacing
``Ottobock 13e200 myobock electrode,'' 2016. [Online]. Available:
  \url{https://professionals.ottobockus.com/c/Electrode/p/13E200~550}
\BIBentrySTDinterwordspacing

\bibitem{HD09}
\BIBentryALTinterwordspacing
P.~Kanerva, ``\BIBforeignlanguage{English}{Hyperdimensional computing: An
  introduction to computing in distributed representation with high-dimensional
  random vectors},'' \emph{\BIBforeignlanguage{English}{Cognitive
  Computation}}, vol.~1, no.~2, pp. 139--159, 2009. [Online]. Available:
  \url{http://dx.doi.org/10.1007/s12559-009-9009-8}
\BIBentrySTDinterwordspacing

\bibitem{HDlowSNR}
A.~Rahimi, S.~Datta, D.~Kleyko, E.~P. Frady, B.~Olshausen, P.~Kanerva, and
  J.~M. Rabaey, ``High-dimensional computing as a nanoscalable paradigm,''
  \emph{IEEE Transactions on Circuits and Systems I: Regular Papers}, vol.~64,
  no.~9, pp. 2508--2521, Sept 2017.

\bibitem{rahimi2016hyperdimensional}
A.~Rahimi, S.~Benatti, P.~Kanerva, L.~Benini, and J.~M. Rabaey,
  ``Hyperdimensional biosignal processing: A case study for {EMG}-based hand
  gesture recognition,'' in \emph{Rebooting Computing (ICRC), IEEE
  International Conference on}.\hskip 1em plus 0.5em minus 0.4em\relax IEEE,
  2016, pp. 1--8.

\bibitem{HD_EEG}
\BIBentryALTinterwordspacing
A.~Rahimi, A.~Tchouprina, P.~Kanerva, J.~d.~R. Mill{\'a}n, and J.~M. Rabaey,
  ``Hyperdimensional computing for blind and one-shot classification of {EEG}
  error-related potentials,'' \emph{Mobile Networks and Applications}, Oct
  2017. [Online]. Available: \url{https://doi.org/10.1007/s11036-017-0942-6}
\BIBentrySTDinterwordspacing

\bibitem{omni}
A.~Moin, G.~Alexandrov, B.~C. Johnson, I.~Izyumin, F.~Burghardt, K.~Shah,
  S.~Pannu, E.~Alon, R.~Muller, and J.~M. Rabaey, ``Powering and communication
  for omni: A distributed and modular closed-loop neuromodulation device,'' in
  \emph{Engineering in Medicine and Biology Society (EMBC), 2016 IEEE 38th
  Annual International Conference of the}.\hskip 1em plus 0.5em minus
  0.4em\relax IEEE, 2016, pp. 4471--4474.

\bibitem{wand}
A.~Zhou, S.~R. Santacruz, B.~C. Johnson, G.~Alexandrov, A.~Moin, F.~L.
  Burghardt, J.~M. Rabaey, J.~M. Carmena, and R.~Muller, ``Wand: A 128-channel,
  closed-loop, wireless artifact-free neuromodulation device,'' \emph{arXiv
  preprint arXiv:1708.00556}, 2017.

\bibitem{nmic}
B.~C. Johnson, S.~Gambini, I.~Izyumin, A.~Moin, A.~Zhou, G.~Alexandrov, S.~R.
  Santacruz, J.~M. Rabaey, J.~M. Carmena, and R.~Muller, ``An implantable
  700$\mu$w 64-channel neuromodulation ic for simultaneous recording and
  stimulation with rapid artifact recovery,'' in \emph{VLSI Circuits, 2017
  Symposium on}.\hskip 1em plus 0.5em minus 0.4em\relax IEEE, 2017, pp.
  C48--C49.

\end{thebibliography}
%



\end{document}